\documentclass[3p,12pt]{elsarticle}
\newcommand{\mysize}{0.87}
\newcommand{\mysizetwo}{0.67}
\usepackage[utf8]{inputenc}
\usepackage{amsmath}
\usepackage{amssymb}
\usepackage[ruled]{algorithm2e}
\usepackage{url}
\usepackage{lineno}
\usepackage{soul}

\usepackage{nomencl}
\makenomenclature

\usepackage{etoolbox}
\renewcommand\nomgroup[1]{%
  \item[\bfseries
  \ifstrequal{#1}{S}{Sets}{%
  \ifstrequal{#1}{P}{Parameters}{%
  \ifstrequal{#1}{V}{Variables}{}}}%
]}

\begin{document}

\nomenclature[S]{$P$}{Set of resource}
\nomenclature[S]{$S$}{Set of supply technologies}
\nomenclature[S]{$Q$}{Set of supply energy type}
\nomenclature[S]{$St$}{Set of storage technologies}
\nomenclature[S]{$M$}{Set of state space models}
\nomenclature[S]{$D$}{Set of demands}
\nomenclature[S]{$W$}{Set of weights}
\nomenclature[S]{$Ex$}{Set of external disturbances}
\nomenclature[S]{$Ext$}{Set of disturbance types}
\nomenclature[V]{$u^{sup}$}{Power generated}
\nomenclature[V]{$v^{sup}$}{Resource consumed}
\nomenclature[P]{$\eta$}{Technology efficiency}
\nomenclature[P]{$tech^{max},tech^{min}$}{Supply limits}
\nomenclature[P]{$Pmax,Pmin$}{Resource consumption limits}
\nomenclature[V]{$SoC$}{Charge level of storage}
\nomenclature[P]{$SoC^{init}$}{Initial storage level}
\nomenclature[P]{$SoC^{max},SoC^{min}$}{Storage level limits}
\nomenclature[V]{$v^{sto}$}{Store charging power/heat}
\nomenclature[V]{$u^{sto}$}{Store discharging power/heat}
\nomenclature[P]{$u^{sto_{MIN}},u^{sto_{MAX}}$}{Storage discharging power/heat limits}
\nomenclature[P]{$v^{sto_{MIN}},v^{sto_{MAX}}$}{Storage charging power/heat limits}
\nomenclature[P]{$\eta^{stl}$}{Continuous storage efficiency}
\nomenclature[P]{$\eta^{stv}$}{Store charging efficiency}
\nomenclature[P]{$\eta^{stu}$}{Store discharging efficiency}
\nomenclature[V]{$u^{in},x,y$}{Demand model input, state and output vectors}
\nomenclature[P]{$A,B,C,E$}{Demand model parameter matrices}
\nomenclature[P]{$d$}{Demand model disturbance}
\nomenclature[V]{$\epsilon$}{Demand model slack variable}
\nomenclature[P]{$sp^{up},sp^{lo}$}{Demand model set-point bounds}
\nomenclature[P]{$dem$}{Demand power/heat requirement}
\nomenclature[V]{$J^{nrg},J^{pen},J^{slack}$}{Objective penalties}
\nomenclature[P]{$\alpha,\beta,\gamma,\phi$}{Objective weighting coefficients}
\nomenclature[V]{$J$}{Total objective cost}
\nomenclature[V]{$Psub$}{Subsystem resource requirement}
\nomenclature[V]{$Ptot$}{Total resource requirement}
\nomenclature[V]{$Pex$}{Predicted resource excess}
\nomenclature[P]{$Plim$}{Total system resource limit}

\title{Integration of an Energy Management Tool and Digital Twin for Coordination and Control of Multi-vector Smart Energy Systems}

\author[label1]{Edward~O'Dwyer\corref{cor1}}
\ead{e.odwyer@imperial.ac.uk}
\cortext[cor1]{Corresponding author}

\author[label2,label3]{Indranil~Pan}

\author[label4]{Richard~Charlesworth}

\author[label5]{Sarah~Butler}

\author[label1]{Nilay~Shah}

\address[label1]{Centre for Process Systems Engineering and Department of Chemical Engineering, Imperial College London, SW7 2AZ, UK}

\address[label2]{Centre for Process Systems Engineering and Centre for Environmental Policy, Imperial College London, SW7 2AZ, UK}

\address[label3]{The Alan Turing Institute, The British Library, London, NW1 2DB, UK}

\address[label4]{Siemens Energy Management, Digital Grid, Nottingham, UK}

\address[label5]{The Royal Borough of Greenwich, The Woolwich Centre, 35 Wellington Street, Woolwich, SE18 6HQ, UK}

\begin{abstract}
As Internet of Things (IoT) technologies enable greater communication between energy assets in smart cities, the operational coordination of various energy networks in a city or district becomes more viable. Suitable tools are needed that can harness advanced control and machine learning techniques to achieve environmental, economic and resilience objectives. In this paper, an energy management tool is presented that can offer optimal control, scheduling, forecasting and coordination services to energy assets across a district, enabling optimal decisions under user-defined objectives. The tool presented here can coordinate different sub-systems in a district to avoid the violation of high-level system constraints and is designed in a generic fashion to enable transferable use across different energy sectors. The work demonstrates the potential for a single open-source optimisation framework to be applied across multiple energy vectors, providing local government the opportunity to manage different assets in a coordinated fashion. This is shown through case studies that integrate low-carbon communal heating for social housing with electric vehicle charge-point management to achieve high-level system constraints and local government objectives in the borough of Greenwich, London. The paper illustrates the theoretical methodology, the software architecture and the digital twin-based testing environment underpinning the proposed approach.
\end{abstract}


\maketitle

\printnomenclature

\section{Introduction}
Cities account for approximately 60-80\% of global energy consumption and are responsible for a similar share of global $CO_2$ emissions \cite{EuropeanComm}. Driven by environmental concerns and urbanisation trends \cite{KeirsteadShah2013}, a clear motivation to significantly reduce the carbon intensity of the urban energy sector has emerged. When considering the transition from fossil fuel dependence to a low carbon future, the presence of large concentrated heating, cooling and electrical power demands in the urban environment poses a significant technical challenge. As the electrification of heating and transport sectors is projected to increase \cite{HMGovernment2017} for example, the infrastructural requirements for a resilient power grid needs to be prioritised. Conversely, this multi-vector demand problem also presents opportunities for cross-cutting synergies to be exploited \cite{Lund2015}. The proliferation of wireless sensing and communication networks and the more widespread adoption of Internet of Things (IoT) concepts provide the opportunity to reduce the environmental impacts of the energy sector while improving system resilience \cite{ODwyer2019}. Key to achieving this is the suitable deployment of computational intelligence and advanced control strategies that can provide optimal operational energy management decisions in real time \cite{Bibri2017}.

The use of Machine Learning (ML) to inform operational actions in the urban energy landscape offers a means of turning ubiquitous data into useful information and distil it to actionable knowledge. Of particular relevance to the energy management problem is the use of ML to infer predictions of future energy consumption from historical data. Examples include Artificial Neural Network (ANN) and Genetic Algorithm (GA) based scheduling approaches \cite{Reynolds2019,Cox2019} and Long Short-Term Memory (LSTM) recurrent neural network models \cite{Wang2019}. The use of advanced control strategies such as Model Predictive Control (MPC) also has well reported advantages over more traditional approaches for energy system management \cite{Killian2016a}. Cost, comfort and environmental objectives to be considered within a framework that can naturally incorporate system limitations and state-feedback. As such, MPC has been applied to smart building control \cite{Kylili2015,Yang2013a}, micro-grid control \cite{Marzband2017} and district-level energy management \cite{Shan2019,Lv2019}.

To enact these strategies, IoT-enabled smart energy management has been investigated in different sectors, with particularly active research in areas such as security and privacy \cite{Braun2018} and communication technologies \cite{Silva2018}. In \cite{Pan2015} for example, an IoT network and control design was developed for energy management in a smart building context, promoting the concept of an intelligent home space while reducing energy consumption. An architecture for a building energy management IoT platform is developed in \cite{Al-Ali2017}, utilising off-the-shelf data analytic and intelligence software to meet consumer demand. At the multi-vector district scale, simulation platforms such as MESCOS \cite{Molitor2014} and IDEAS \cite{Baetens2012} have been proposed to capture the key interactions between interconnected energy systems incorporating various supplies and demands.

As new methods are developed and refined, energy management approaches must evolve appropriately and, to this end, a single reconfigurable framework that can contain forecasting and optimisation capabilities for real-time energy management is still required. While applications can vary across different sectors, many common underlying requirements exist, including optimisation formulations and solvers, model structures, data acquisition approaches and forecasting methods. To flexibly allow for different energy system types and contexts while enabling the incorporation of new functions and capabilities, an extensible framework would be required in a modular form to enable different configurations and user-preferences to be incorporated without significant software re-development. The goal of this paper is the development of an open-source software framework for use by local government to coordinate interconnected energy assets in a city or district. This tool, denoted the Sustainable Energy Management System (SEMS), is a set of modules incorporating advanced control in the form of MPC as well as ML-based forecasting algorithms to provide optimal operational decisions to the energy networks of a smart city or district.

The SEMS decomposes the overall district-level problem into a set of subsystems and derives an optimal operational trajectory for each. A high-level coordination module then evaluates the proposed solutions from the district-level perspective, enforcing mitigating action in times of constraint violation. The SEMS consumes measurements and information relating to external influences in real time, sending updated set-points following a receding horizon-based MPC approach and would be used by owners of multiple flexible energy assets such as local government bodies. 
Given the complexity of the decision-making tasks involved, a key feature of the proposed methodology is the use of a suitable digital twin environment to test different approaches and scenarios, enabling analysis to be carried out from different stakeholder perspectives. The SEMS is integrated with an open-source digital twin tool to achieve this to be used both prior to implementation for design and in parallel with real-world operation to allow adaptive evaluation and updating of strategies.

The main contributions of the work are as follows. The framework can be applied to muti-vector energy systems incorporating different applications while including a coordination layer to ensure the various sub-systems adhere to higher-level constraints. This is particularly important given the current push for greater electrification of heating and transport sectors. The extensible framework proposed allows different prediction algorithms and optimisation formulations to be implemented out without a major overhaul of the underlying code. Though a range of scenarios are presented in the paper, the SEMS has not been developed to fit to one application only, but to allow for scaling and transferal to different applications while incorporating different forecasting methods. Finally, the integration of the SEMS with a digital-twin simulation environment allows for potentially complex decisions to be informed by scenario-based analysis, thus ensuring that problems can be framed in a more tangible context. All components of the tool are developed with open-source software to encourage replication and growth.

In Section \ref{Modules}, a detailed description of the different modules of the SEMS is presented, while the integration and deployment of the SEMS is then discussed in Section \ref{SimMod}. Finally, the application of the SEMS to the EU-funded Sharing Cities project is presented. This includes a set of case studies used to illustrate the ability of the SEMS to control and coordinate assets across different energy sectors. 

\section{Modular control and scheduling framework}\label{Modules}
\subsection{Requirements for flexibility and transferability}
A wide range of possible applications can exist in the energy management of a district, and no two district or cities will be the same. A tool that is over-specified to one context will not be transferable to others. As such, the SEMS developed here is designed to be generic and modular, thus encouraging adaptation to different contexts. In this generic form, each subsystem is considered as a network of elements representing supplies, demands and energy stores generating useful energy from multiple resources to satisfy economic, environmental and user-centric objectives. The demand elements can be in the form of static time-series (if a demand cannot be influenced by the SEMS), or in the form of a state-space model (commonly used in MPC problems). In current form, the state-space model represents the dynamic relationship between the demand inputs and outputs - the SEMS will seek to choose a set of energy inputs that provide some desired behaviour in the demand outputs (e.g. maintaining the temperature in a building within comfort bounds). Other model forms could be readily incorporated, including physics-based approaches and could be in a discrete-event or hybrid switched mode form. By leveraging pre-existing open-source tools, best-practice approaches can be developed to fit specific applications. An arbitrary number of energy vectors can be included, and the energy supplies and stores can correspond to any class of technology. 

To further encourage the transferability of the SEMS, it is composed of solely open-source software, avoiding costs and barriers associated with licensing (the SEMS core itself is written in Python3.6). It is envisaged as an ever-evolving set of interchangeable modules which can expand naturally as new applications arise, rather than a single strictly defined pre-packaged set of algorithms. The following sections outline the modules currently used for model identification, forecast generation, control and coordination. However, if an alternative algorithm for one of these modules is preferred, it should be possible (within reason) to replace an existing algorithm with another, while keeping the rest of the framework intact (due to the modular architecture as discussed before). For the general form of the SEMS, robust linear approaches are then favoured over more sophisticated but application-specific non-linear approaches. The purpose of the SEMS is to deliver set-point recommendations to a lower-level control/actuator layer (for the sake of robustness and stability, it would not be desirable to directly send control signals to the assets using a cloud-based solution such as this \cite{Maciejowski}).

\subsection{The SEMS subsystem modules}\label{SecSubProblem}
\begin{figure*}
\centering
\includegraphics[width=\mysize\textwidth]{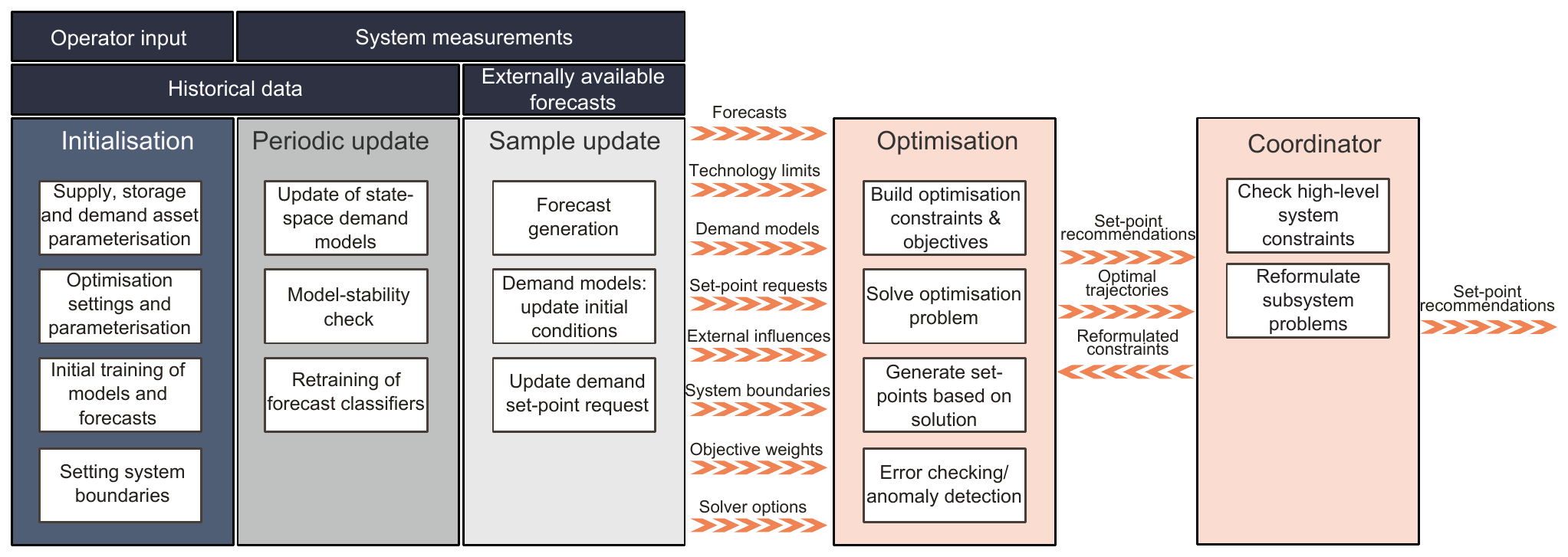}
\caption{\label{fig:SEMSMod}Overview of individual sub-system optimisation}
\end{figure*}

A schematic of the algorithm structure employed by the SEMS for each subsystem is shown in Fig. \ref{fig:SEMSMod}. The problem is first defined in terms of the system component parameters, boundaries and energy vectors as well as the modelling, forecasting and optimisation options to be used. In long-term intervals, re-parameterisation can be carried out to capture any slow or seasonal trends. Model states, set-points and generated forecasts are then updated at each time sample. The optimisation problem is solved and the optimally derived set-points, inputs and outputs for the following day are sent to the high-level system coordinator. If the high-level constraints are satisfied, the first elements of the optimal set-point trajectories are then sent to the real system. Otherwise, the system coordinator reformulates the subproblem constraints following the approach described in Section \ref{Coordinate}.

The specific components that form this general approach are defined in detail in the following sections.

\subsubsection{Forecast generation}\label{MLForecast}
Data-driven and ML approaches have become more common for forecasting temporal variables based on historical measurements, particularly when the underlying physical dynamics of the systems described by the variables are analytically complex. Examples include Artificial Neural Network (ANN) approaches for renewable energy generation \cite{Reynolds2019} and building electricity demands \cite{Chae2016}. To give a user the option of such a forecasting service, modules based on supervised ML techniques available using the Scikit-learn toolbox \cite{ScikitLearn} are included in the SEMS framework.

To illustrate the use of forecasting within the overall SEMS framework, two possible algorithms are described here. This is an active research area with many different routes for ML-based forecast generation possible and different approaches will be more applicable to different situations (a review is presented in \cite{Ahmad2018} for example). As such, it must be emphasised that the algorithms in this section should only be seen as place-holders that can be replaced by different algorithms depending on the application and user preference insofar as they can fit within the generic framework, consuming data from the data-store and passing forecasts to the optimisation module. The first of the two approaches used is based on classification and clustering and is suited to forecasting demand or profiles with a repetitive profile (diurnal energy demands for example), while the second is a more typical regression-based approach using ANNs.

An overview of the methods is provided here, with examples of the application of the algorithms to energy system forecasting problems presented in Section \ref{results}. It should be noted that the solvers and algorithms were chosen based on their performance in the applications presented - as stated, other configurations may be better suited to different applications. 

\begin{figure*}
\centering
\includegraphics[width=\mysizetwo\textwidth]{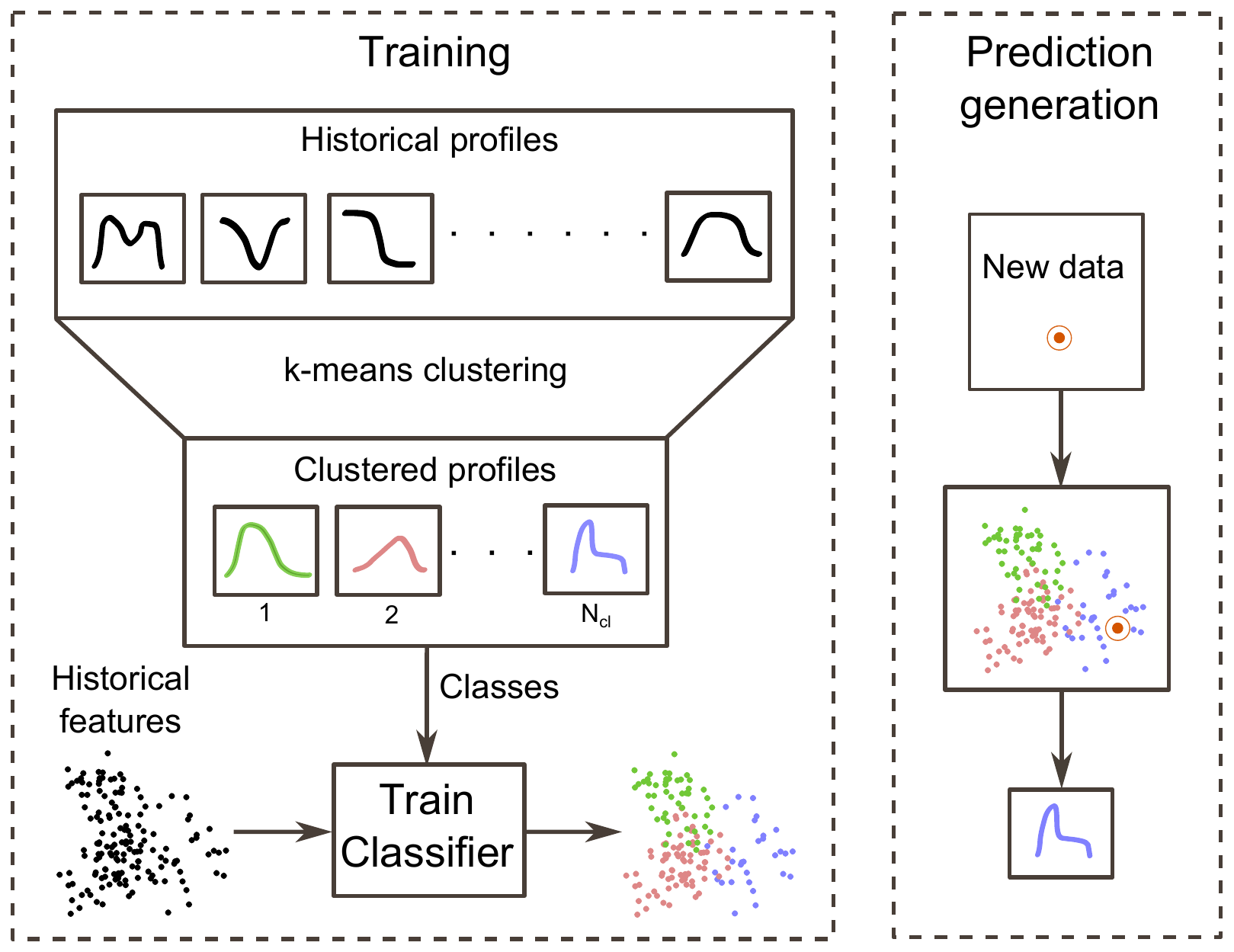}\caption{\label{fig:ClassSVM}Classification-based ML approach for forecast generation}
\end{figure*}

The concept behind the first approach centres on clustering and classification. The training data is broken into daily chunks, and a k-means clustering algorithm is used to develop a set of representative cluster prototypes. Using an ensemble classification approach such as Random Forest (RF) or Gradient Boost (GB), the forthcoming day is assigned to its most representative cluster prototype. This approach exploits the daily patterns that emerge in many energy system applications and ensemble approaches are reasonably robust to over-fitting (which is particularly useful if the data quality is low). An example of such an approach would be the prediction of a building heating demand profile. The training feature set could comprise day-type information (e.g. weekend, weekday, holiday) and external temperature information. This is similar to the time series aggregation approaches developed for seasonal storage operation in \cite{Kotzur2018} and \cite{VanderHeijde2019}. A graphical representation of the approach is given in Fig.\ref{fig:ClassSVM}.

Alternatively, if common daily patterns are not expected, a regression-based ANN approach can be used. Conceptually, this approach is more straightforward. Using a standard multi-layer perceptron model and stochastic gradient descent optimisation algorithms, neural network weights can be derived with an input training data-set. The resulting relationship between features and outputs is then used to project future outputs for the forthcoming day. Different neural network architectures might work better in many instances, for example recurrent neural network and LSTM models have often been shown to perform well in time series predictions like this. The modelling paradigm is open for inclusion of all these ML model variants. 

\subsubsection{Control formulation}
The primary aim of an energy management tool is to provide control decisions to the participating assets, for applications such as temperature regulation and supply technology choices in heat networks, energy storage deployment, electric vehicle charger curtailment for grid resilience etc. MPC, an advanced control technique, is a natural fit for such applications to its ability to optimise control actions to a given set of criteria while incorporating future system behaviour and explicitly handling system constraints. Applications are often limited to academic study, partly due to the required implementation effort \cite{Sturzenegger2016a}. Implementing MPC at this scale introduces challenges in terms of handling the complexity of interconnected systems, modelling the relevant dynamics and implementing a software package that does not rely on proprietary algorithms and architectures. The proposed approach seeks to overcome many of these barriers, significantly expediting the design effort. The following features ensure that an appropriate condition is provided. A pre-defined constraint formulation that incorporates energy supply, storage and demand components is formed. This can be expanded or reduced as desired to fit to new applications and scenarios. Objective formulations are provided allowing for energy, cost, $CO_{2}$ and/or discomfort minimisation, these being of particular interest in the energy domain. This optimisation formulation, at the heart of the MPC problem, is designed to be solved with open-source solvers and adaptable to different problems with little adjustment. Perhaps most importantly, as part of the framework, multiple MPC subproblems can be formed and coordinated via a high-level master coordinator to account for the overlapping constraints and objectives of the district-level energy landscape. The subproblems can integrate with forecasting tools and external data repositories to allow for external factors to be considered in the optimisation problem without the compatibility issues that can arise when combining software tools. The framework can also be integrated with an open-source simulation tool for analysis and evaluation of strategies that may cut across multiple energy vectors. This can significantly reduce the time needed for the modelling stage of the control design process. These features are described in detail in the following sections.

\paragraph{State-space model generation:}
Formulating demand models as linear state-space systems has advantages for the definition of the MPC problem. It may be necessary to parameterise these state-space models using data-driven techniques. The SEMS once again employs the Scikit-learn library to do this, in this case using a ridge regression algorithm to choose model parameters that minimise the difference between historical measurements and modelled predictions of these historical terms. This approach has been included to illustrate the modelling ability, but it could be greatly expanded upon. For example, in applications containing non-linear dynamics, models would be linearised around particular operating points with the potential to switch between multiple models as the system moves from one operating point to another. Such approaches could be incorporated without affecting the structure of the tool.

\paragraph{Optimisation Constraints:}
In this section, the constraints of the optimisation at the core of the MPC formulation associated with an individual subsystem are presented. In the SEMS, the problem is formulated using the Pyomo library \cite{PyomoPaper,PyomoBook}.

A supply technology (e.g. a boiler, PV panel, heat pump etc.) (\(s\in S\)) converts energy from a resource or multiple resources to useful energy of type or multiple types, with a conversion efficiency from resource \(p\in P\) to energy type \(q\in Q\) given by \(\eta_{s,p,q}\). Where the superscript $sup$ is used throughout to denote variables relating to the supply technologies, the relationship between the energy generated (\( u^{sup}_{s,q,i}\)) and resource consumed (\(v^{sup}_{s,p,q,i}\)) is given for each step along the prediction horizon (\(i\in [0,1, ... ,N-1]\)) by the following:
\begin{gather}\label{eq:firsteq}
  u^{sup}_{s,q,i} = \sum_{p\in P}v^{sup}_{s,p,q,i}\eta_{s,p,q}\\
  \forall s\in S,\forall q\in Q,\forall i\in[0,1, ... ,N-1] 
\end{gather}
The supply limits are constrained between upper and lower bounds, denoted \(tech^{max}\) and \(tech^{min}\) respectively:
\begin{gather}
  v^{sup}_{s,p,q,i}\leq tech^{max}_{s,p,q,i}\\
  v^{sup}_{s,p,q,i}\geq tech^{min}_{s,p,q,i}\\
  \forall s\in S,\forall p\in P,\forall q\in Q,\forall i\in [0,1, ... ,N-1]\nonumber
\end{gather}
Such supply technologies can represent heating, cooling and electrical generation assets in the system (e.g. boilers, Combined Heat and Power (CHPs), heat pumps, PV panels or wind turbines).

The consumption of each energy resource by all supplies is limited between upper (\(Pmax_{p,i}\)) and lower (\(Pmin_{p,i}\)) bounds as follows:
\begin{gather}
  Pmin_{p,i}\leq\sum_{s\in S,q\in Q}v^{sup}_{s,p,q,i}\leq Pmax_{p,i}\\
  \forall p\in P, i\in [0,1, ... ,N-1]\nonumber
\end{gather}
For renewable resources such as solar radiation or wind, the upper limits would correspond to the available power, whereas for a resource such as gas or electricity, the limits may be associated with limits imposed by an energy supplier (or self-imposed).

Storage technologies (\(st\in St\)) in the model are modelled with a charging efficiency (\(\eta^{stv}_{st}\)), discharging efficiency (\(\eta^{stu}_{st}\)) and storage loss efficiency (\(\eta^{stl}_{st}\)). These storage technologies can represent batteries and thermal stores, but also such a technology can represent the delay of the use of an EV charger (from an energy balance perspective, the delayed consumption can be seen as equivalent to a battery with a negative storage capacity). Variables associated with the storage technologies are given the superscript $sto$ throughout, where \(v^{sto}_{st,q,i}\) and \(u^{sto}_{st,q,i}\) are the charging power and discharging power respectively associated with energy type \(q\) at time \(i\), the state of charge (\(SoC_{st,q,i}\)) of the storage technology \(st\) is defined as:
\begin{gather}
  SoC_{st,q,i+1} = \eta^{stl}_{st}SoC_{st,q,i}-\frac{1}{\eta^{stu}_{st}}u^{sto}_{st,q,i}+\eta^{stv}_{st}v^{sto}_{st,q,i}\\
  \forall q\in Q,\forall i \in [0,1, ... ,N-1]\nonumber
\end{gather}
The state of charge is initialised with the measured value (\(SoC^{init}_{st,q}\)), and the upper (\(SoC^{max}_{st,q}\)) and lower bounds (\(SoC^{min}_{st,q}\)) are set using the following constraints: 
\begin{gather}
  SoC_{st,q,0} = SoC^{init}_{st,q}\\
  SoC^{min}_{st,q} \leq SoC_{st,q,i} \leq SoC^{max}_{st,q}\\
  \forall st\in St,\forall q\in Q,\forall i \in [0,1, ... ,N-1]\nonumber
\end{gather}
The stores are limited by upper and lower bounds on the charge and discharge powers as follows: 
\begin{gather}
  u^{sto_{MIN}}_{st,q}\leq u^{sto}_{st,q,i}\leq u^{sto_{MAX}}_{st,q}\\
  v^{sto_{MIN}}_{st,q}\leq v^{sto}_{st,q,i}\leq v^{sto_{MAX}}_{st,q}\\
  \forall st\in St,\forall q\in Q,\forall i \in [0,1, ... ,N-1]\nonumber
\end{gather}
Furthermore, the charging power of a store is constrained to be less than or equal to the power supply generated by the supply assets connected to that store. With the subset of store-connected supplies given as $S^{st}\subseteq S$, this limit is defined by the following constraints:
\begin{gather}
  \sum_{st\in St}v^{sto}_{st,q,i}\leq \sum_{s\in S^{st},p \in P}u^{sup}_{s,p,q,i}\\
  \forall q\in Q,\forall i \in [0,1, ... ,N-1]\nonumber
\end{gather}

For MPC-type problems, energy demands can be characterised in the form of a linear state-space model to which the energy of type \(q\), generated by the supply assets and discharged by the stores, forms the input vector \(u^{in}_{q,i}\) as given by:
\begin{gather}
  u^{in}_{q,i}=\sum_{st\in St}(u^{sto}_{st,q,i}-v^{sto}_{st,q,i})+\sum_{s\in S}u^{sup}_{s,q,i}\\
  \forall q\in Q, \forall i \in [0,1, ... ,N-1]\nonumber
\end{gather}
The demand model can also be influenced by external disturbances (\(ext\in Ext\)) of type \(ex\in Ex\) (in a model representing the thermal behaviour of a building for example, the internal temperature may be influenced by external factors such as solar radiation and ambient temperature). For such a demand element \(m\), the relationship between these inputs and disturbances with the vectors of states (\(\mathbf{x}_{m,i}\)) and outputs (\(\mathbf{y}_{m,i}\)) are described by the following equations, where \(A_{m}\), \(B_{m}\), \(E_{m}\) and \(C_{m}\) represent the coefficient matrices of the model:
\begin{gather}
  \mathbf{x}_{m,i+1}=A_{m}\mathbf{x}_{m,i}+\sum_{q\in Q}B_{m,q}u^{in}_{q,i}\nonumber\\+\sum_{ex\in Ex}E_{m,ex}\sum_{ext\in Ext}d_{ex,ext,i}\\
  \mathbf{y}_{m,i}=C_{m}\mathbf{x}_{m,i}\\
  \mathbf{x}_{m,0}=\mathbf{x}^{init}_{m}\\
  \forall m\in M, \forall i \in [0,1, ... ,N-1]\nonumber
\end{gather}
A soft constraint is used to penalise deviations of the output from the desired set-point in the MPC formulation, while an allowable set-point range can also be included. Where the upper and lower set-point bands are given as \(sp^{up}_{m,i}\) and \(sp^{lo}_{m,i}\) respectively, the slack variables representing the deviation of the output outside the specified band are defined as:
\begin{gather}
  \mathbf{\epsilon}_{m,i}\geq\mathbf{y}_{m,i}-\mathbf{sp}^{up}_{m,i}\\
  \mathbf{\epsilon}_{m,i}\geq\mathbf{sp}^{lo}_{m,i}-\mathbf{y}_{m,i}\\
  \forall m\in M, \forall i \in [0,1, ... ,N-1]\nonumber
\end{gather}

While some demands may fit this aforementioned MPC structure (e.g. building thermal systems), it may be preferable to view other demands in the form of a static time-series as opposed to a dynamic model, particularly if the demand cannot be directly altered. In such cases, the MPC problem becomes a scheduling problem based on energy balances between the supplies, stores and demands and the state space demand model is replaced by a time-series demand forecast. The following constraint is used to ensure that the supply and storage mix chosen matches the total demand requirement (where each individual demand \(de\in D\) is defined for energy type \(q\) at time \(i\) as \(dem_{{de,q,i}}\)):
\begin{gather}
  \sum_{s\in S,p\in P}u^{sup}_{s,q,i}+\sum_{st\in St}(u^{sto}_{st,q,i}-v^{sto}_{st,q,i})=\sum_{de\in D}dem_{de,q,i}\\
  \forall q\in Q, \forall i\in[0,1, ... ,N-1]\nonumber
\end{gather}

\paragraph{Objectives:}
The optimisation objectives (denoted here by $J$) forming the cost function of an individual subsystem are next summarised. Three components are included in the overall objective. Factors associated with energy consumption are denoted by the superscript $nrg$, slack variable costs are denoted by the superscript $slack$ and additional miscellaneous user-specified penalties are denoted by $pen$.

Different factors (denoted \(w\in W\)) associated with the consumption of a resource (e.g. the \(CO_{2}\) emissions and financial cost associated with gas or electricity consumption) can be penalised, weighted by the coefficient \(\alpha_{w,p,i}\in [0,1]\). Larger values (closer to 1) imply that the associated term is more heavily penalised in the objective, while setting it to 0 implies that the term is not considered in the optimisation. The ratio of the weights across all objectives define the balance of the overall cost function, enabling a user to place more or less emphasis on different objectives. The following equation defines the objective corresponding to the energy related terms: 
\begin{gather}
  J^{nrg}_{i} = \sum_{w\in W,p\in P,q\in Q}\alpha_{w,p,i}v^{sup}_{s,p,q,i}\\
  \forall i\in[0,1, ... ,N-1]
\end{gather}
The use of a storage technology can also be penalised using the weighting coefficient \(\beta_{st,i}\in [0,1]\), with larger weights again corresponding to greater penalties in the cost function, as follows:
\begin{gather}
  J^{pen}_{i} = \sum_{q\in Q,st\in St}\beta_{st,i}SoC_{st,q,i}
\end{gather}
The purpose of such a penalty may be, for example, to discourage the delaying of EV chargers or the overuse of batteries.

The presence of slack variables representing deviations of state-space model outputs from their corresponding set-points has been described previously. To penalise these deviations in the cost function, the weighting coefficient \(\gamma_{m}\in[0,1]\) is used in the same manner as the previous weighting coefficients in the following equation:
\begin{gather}
  J^{slack}_{i} = \sum_{m}\gamma_{m}\epsilon_{m,i}^2
\end{gather}
These set-points may be one-sided, for example whereby the goal is to only penalise deviations below a set-point, not above (particularly relevant in heating system applications).  

These weighted objectives are combined in a single cost function. Once again, these must be appropriately weighted to ensure all terms are of the same scale and are balanced as per the user requirements. The weight applied to the energy consumption terms miscellaneous terms and slack variable terms are denoted $\phi_{nrg}\in[0,1]$, $\phi_{pen}\in[0,1]$ and $\phi_{slack}\in[0,1]$ respectively. The following then represents the overall cost function to be minimised for the subsystem:

\begin{gather}\label{eq:lasteq}
  J = \sum_{i=0}^{N-1}\left(\phi_{nrg}J^{nrg}_{i}+\phi_{pen}J^{pen}_{i}+\phi_{slack}J^{slack}_{i}\right)\\
  \phi_{nrg}+\phi_{pen}+\phi_{slack}=1
\end{gather}

The formulation described is based on a deterministic optimisation. Given the uncertain nature of certain energy system components (energy demands may be heavily influenced by non-deterministic user behaviour for example), a stochastic or robust formulation may be appropriate. Although such an approach has not been applied here, it should be emphasised again that the specific formulation can be adapted for different applications in the future. For the problems considered in this work, deterministic optimisation is deemed adequate due to the slow nature of the underlying dynamics and the potential to use the digital twin to simulate a range of possible scenarios representing different levels of uncertainty in the system.

\subsection{Coordination layer for system-level cooperation}\label{Coordinate}
At a district level, the overall energy landscape may be composed of several subsystems, each of which can be handled by the optimisation formulation described in Section \ref{SecSubProblem}. Though these subsystems may be largely isolated from each other, they may draw power from the same energy sources. An example of this is the case whereby multiple heat networks and/or transport networks are dependent on the same electricity network. Coinciding demand peaks across the different subsystems (potentially quite likely due to coordinated patterns of behaviour) can lead to excessive grid stress if the optimisation of each subsystem is only carried out in isolation. As electrification of heating and transport sectors increases, this scenario becomes more relevant. An important feature of the SEMS is the ability to coordinate the different subsystems when the total source requirement is predicted to exceed a specified threshold. 

For a subsystem $\ell$ described by Eqns. (\ref{eq:firsteq}) - (\ref{eq:lasteq}), the predicted resource (of type \(p\)) requirement at time \(i\) (\(Psub_{\ell,p,i}\)) can be defined as:
\begin{gather}\label{eq:sub}
  Psub^{\ell}_{p,i}=\sum_{s\in S,q\in Q}v^{sup}_{s,p,q,i}
\end{gather}
For $L$ subsystems, the total predicted requirement of resource $p$ over the prediction horizon is then:
\begin{gather}\label{eq:tot}
  Ptot_{p,i}=\sum_{\ell=1}^{L}Psub_{\ell,p,i}, \forall i\in [0,1, ... ,N-1]\\
  \forall p\in P, i\in[0,1, ... ,N-1]\nonumber
\end{gather}
Where the system-wide resource limit is denoted \(Plim_{p,i}\), the excess resource requirement is then defined as $Pex_{p,i}$, given as:
\begin{gather}\label{eq:excess}
  Pex_{p,i} = max(0,Plim_{p,i}-Ptot_{p,i})\\
  \forall p\in P, i\in [0,1, ... ,N-1]
\end{gather}

If this excess resource consumption is greater than 0 for any resource, the limits associated with that resource are tightened in each subsystem sequentially until the system-wide resource limit is satisfied. The subsystems must be set in a prioritised order \(1...L\) such that \(\ell=1\) corresponds to the subsystem most open to resource restriction and \(\ell=L\) corresponds to subsystem in which resource restriction is least desirable. The coordination algorithm is described in Algorithm \ref{Alg}. 

\begin{algorithm}
\SetAlgoLined
\KwResult{Predicted total power demand is less than system capacity threshold}
 $\ell\leftarrow0$\;
 $Pex_{p,i} = max(0,Plim_{p,i}-Ptot_{p,i}), \forall p,\forall i$\;
 \While{$Pex_{p,i}\geq 0, \forall p,\forall i$ \bf{and} $\ell\leq L$}{
 $\ell\leftarrow\ell+1$\;
 \For{$p\in P$}{\For{$i\leftarrow0$ \KwTo $N-1$}{$\bar{P}max^{\ell}_{p,i}\leftarrow\max(Pmin^{\ell}_{p,i},Psub^{\ell}_{p,i}-Pex_{p,i})$\;
 $Pmax^{\ell}_{p,i}\leftarrow\bar{P}max^{\ell}_{p,i} \forall p,\forall i$\;}}
 reformulate and solve subproblem $\ell$\;
 update $Psub_{l,p,i}$ and $Ptot_{p,i}$ using Eqns. (\ref{eq:sub}) and (\ref{eq:tot})\;
 $Pex_{p,i}\leftarrow Plim_{p,i}-Ptot_{p,i}, \forall p,\forall i$\;
 }
 \caption{\label{Alg}Coordination of multiple sub-systems}
\end{algorithm}

\begin{figure*}
\centering
\includegraphics[width=\mysizetwo\textwidth]{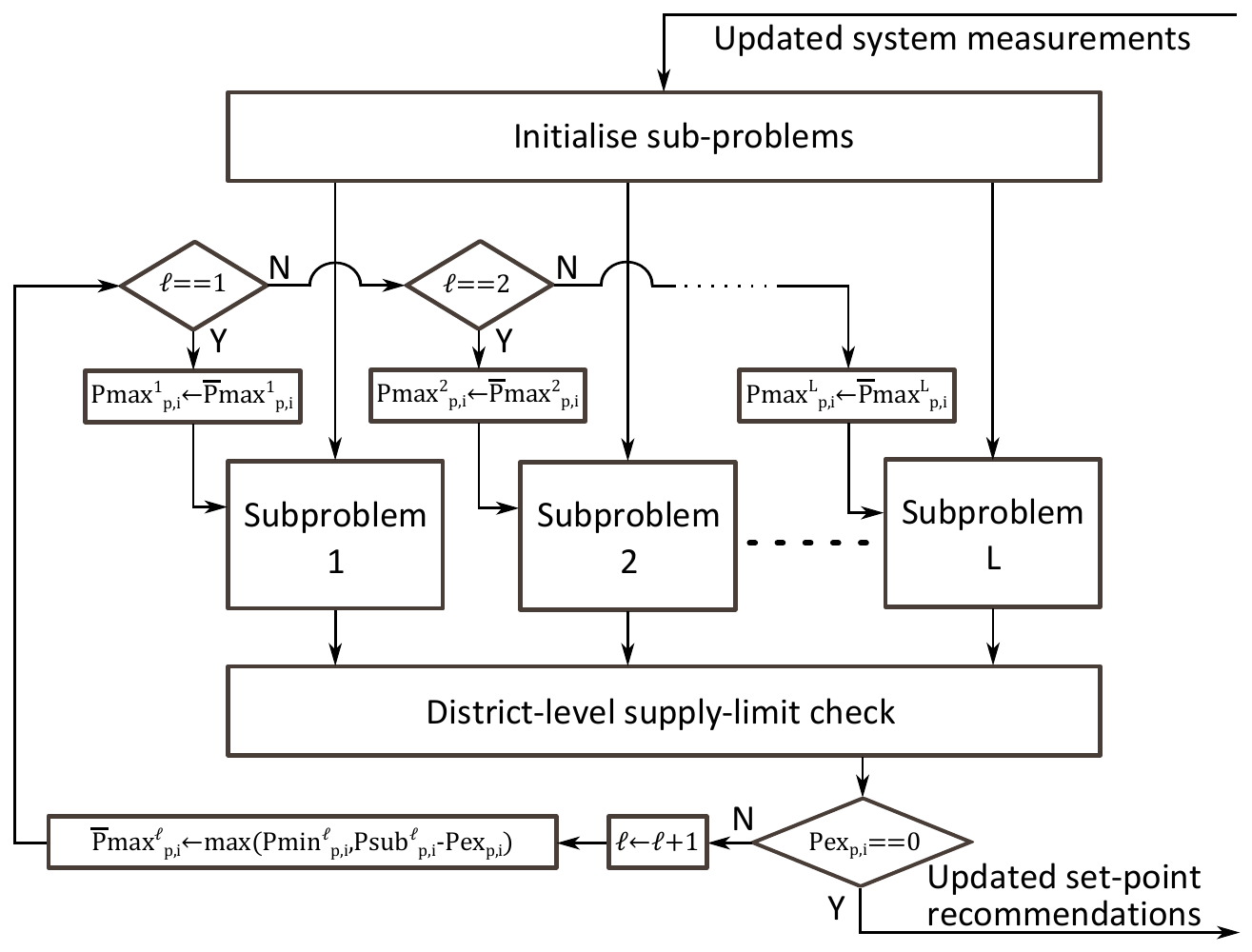}
\caption{\label{fig:AlgDiag}Coordination of subsystems for district level supply constraint satisfaction}
\end{figure*}

A graphical representation of the coordination mechanism is shown in Fig.\ref{fig:AlgDiag}.

\section{Integration of the SEMS and system simulation}\label{SimMod}

\subsection{Simulation modelling platforms for development and analysis}
At the district or neighbourhood level, energy systems can be composed of many possible subsystems in different configurations, while stakeholder priorities can be diverse. As previously discussed, where new algorithms and coordination approaches are to be applied, there is a need to test and refine strategies \textit{in silico} prior to installation. 

Acting as a digital surrogate for the real system, a digital twin can combine on-line measurement updates with simulated system behaviour to improve operational decision-making strategies while providing more detailed performance insight and aiding the identification of faults and anomalies in the system. Furthermore, uncertainties in the models and predictions can be analysed through appropriate scenario testing. Underpinning a digital twin is a simulation environment capable of replicating the behaviour of the real system. Such a simulation environment is a vital tool in the development of a district-level energy management scheme. Through off-line simulations, software interfaces can be tested, control algorithms and operational strategies can be rigorously evaluated and proposed technological interventions can be validated. As part of the development of the energy management system, a suitable simulation environment is presented here. 

While simulation of the various sub-systems within an urban energy setting for these purposes is commonplace \cite{Allegrini2015}, the multi-scale, interconnected nature of the challenges faced, poses a problem for many common modelling software packages. The need to apply advanced control and ML-based strategies across various components while considering the interactions between the systems implies that a common modelling framework is required that can capture the various time-scales present in a multi-vector energy system, constructed in a modular format to ensure transferability and scalability. To satisfy these requirements, a simulation environment was developed using the Ptolemy II open-source modelling framework \cite{PtolemyII} using the Building Controls Virtual Test-bed (BCVTB) environment \cite{BCVTB}. This allows for concurrent co-simulation of different software packages including EnergyPlus \cite{Eplus}, which is used here as the building energy modelling tool. Once again, only open-source tools were included to encourage replication.

\subsection{Simulation modules and high-level architecture}
\begin{figure*}
\centering
\includegraphics[width=\mysize\textwidth]{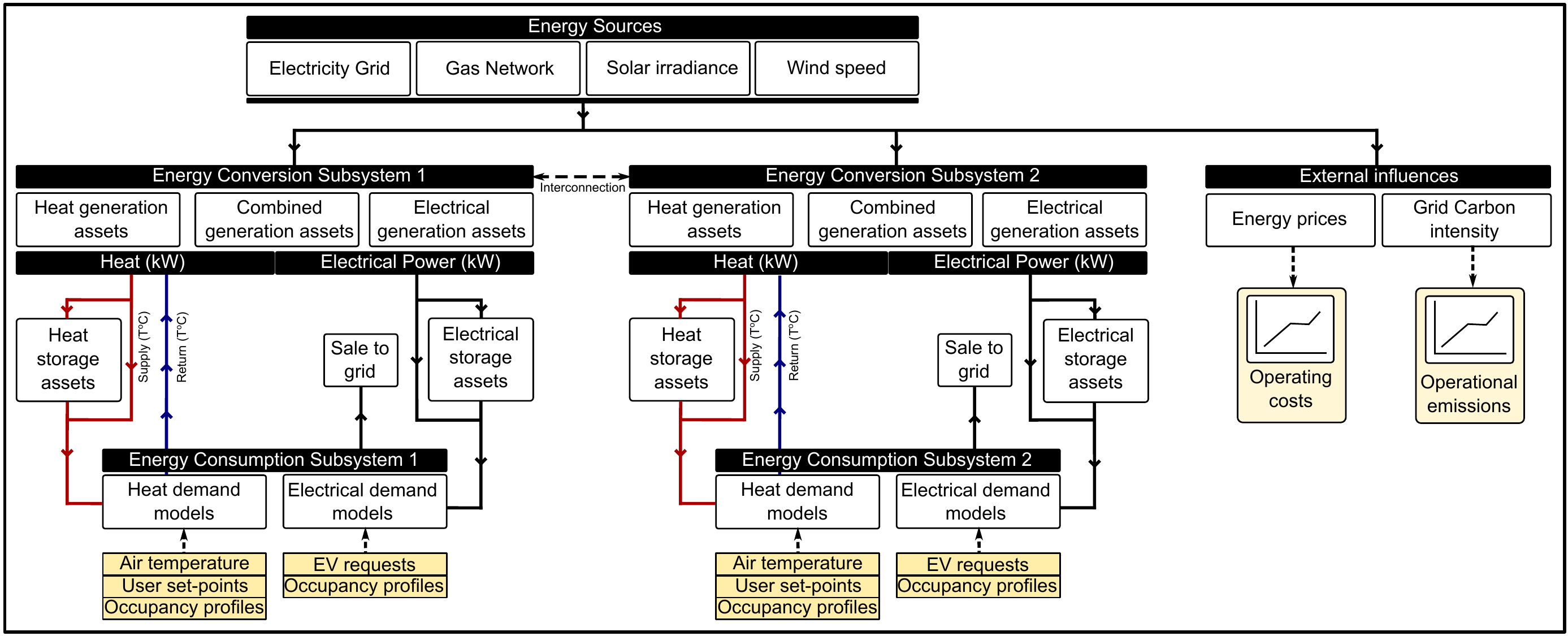}
\caption{\label{fig:TwinSim}High-level component architecture of simulation environment with two sub-systems}
\end{figure*}

The simulation environment can be viewed as a combination of energy source characteristics, conversion and storage technology models, demand models and external influence profiles (weather, energy prices, emission indicators etc.). Energy sources include gas, the electricity grid and renewable sources such as wind and solar radiation. Conversion technologies act as a bridge between the energy sources and the demands and can include heat generation (e.g. boilers, heat pumps), power generation (e.g. PV panels), cooling (e.g. chillers) or combined generation (e.g. CHPs), while storage technologies can be thermal or electrical in nature. Energy is conveyed from the conversion technologies to the demands (heat transfer is modelled in terms of temperatures and flow rates), managed via valves, switches and actuators and low-level control laws (e.g. PI controllers). The demand models characterise the consumption of this energy. Building heat demand is modelled using EnergyPlus, a software package in which detailed knowledge of the physical building envelope is used to generate a representative model of the heat requirement. The high-level architecture for the simulation environment is shown for a system with two generic subsystems in Fig. \ref{fig:TwinSim}.

Key to the validity of the simulation environment is the ability to communicate with the energy management platform in a manner that closely replicates the real system. To achieve this link, the Flask \cite{Flask2014} web-app framework is used to pass measurements from the digital representation of the system to the SEMS and to return the derived set-points from the SEMS back to the twin. 

\subsection{Integrating with the real-world}

\begin{figure*}
\centering
\includegraphics[width=\mysize\textwidth]{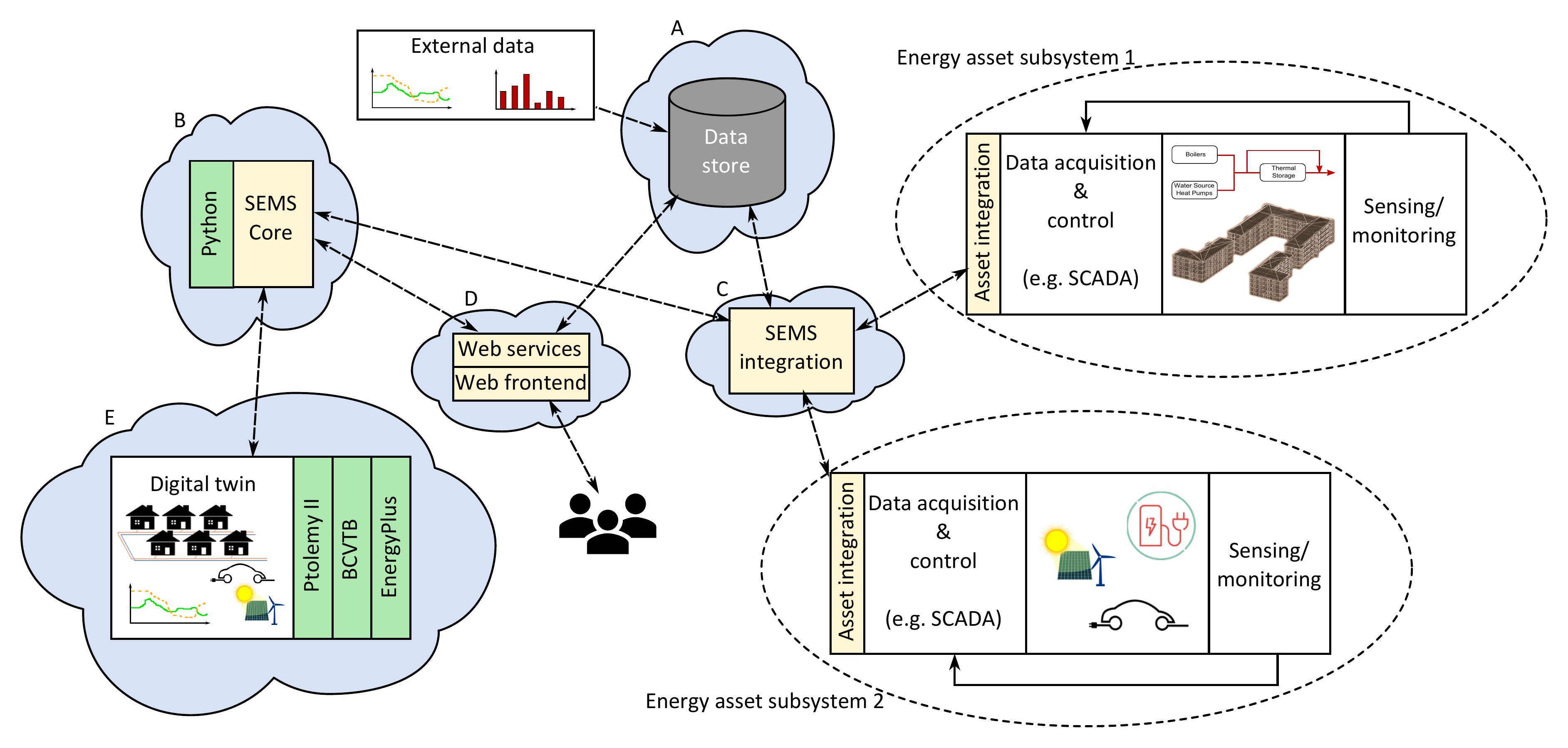}
\caption{\label{fig:TwinSEMS}Data flows between the SEMS, the system components and the digital twin}
\end{figure*}

In Fig. \ref{fig:TwinSEMS} the communication architecture between the simulation environment and the energy management system is shown. 

The overall system consists of real physical assets with controllers (multiple sub-systems encircled with black dashed boundaries) interfaced with cloud based services indicated in blue (databases, SEMS algorithmic modules and digital twin). Each cloud service is containerised, hosted independently and communicates with the other modules through API calls. This ensures modularity, rolling updates for each service without updating the full system and resilience (multiple similar containers can be deployed on the cloud for reliability and load balancing purposes). 

The centralised data store service (Label A) can take inputs about system variables (from energy asset sub-systems) and environmental variables (from external sources, either modelled or in real time). These data-sets are queried by the SEMS core algorithms (Label B) with a web services /front-end interface (Label D) between the two. The web frontend allows the display of the system states and the controller actions through easily understandable graphs. The SEMS core block computes the set points and pushes them to the data-store. These updated set-points are pushed to the low-level control system of each asset through the SEMS integration module (Label C). The SEMS core makes decisions after updating the digital twin (Label E) with real time data, constraints etc. and receiving predictions from the digital twin. 

\section{Implementation: The SEMS in the Sharing Cities project}
\subsection{Application to the smart city initiatives of Greenwich}\label{GreenwichSetup}
To demonstrate the use of the SEMS tool, the approach was applied to the smart cities interventions currently underway in the borough of Greenwich in London as part of the EU-funded Sharing Cities project \cite{SharingCities}. The proposed activities include interventions in the built environment through retrofit and heating system electrification in part of the social housing stock and interventions in the transport sector through increased EV charger installation as well as the introduction of additional renewable generation capacity in the form of PV panels. The push for electrification of heating and transport sectors leads to a multi-vector, integrated energy landscape, in which electrical system capacity may be stretched without proper energy management approaches that can handle the interconnected nature of the various assets. In this way, it is highly suited to the deployment of an intelligent energy management system such as the SEMS.

The development of a simulation environment that can represent the relevant energy systems in Greenwich is outlined in detail in \cite{ODwyer2019a}. Using the Ptolemy II-based platform described in Section \ref{SimMod}, two subsystem models were developed. The first represents the proposed upgraded heating system of the Ernest Dence estate, a social housing complex comprising 95 flats with annual heating energy usage in its present form of approximately 2 GWh.

\begin{figure*}
\centering
\includegraphics[width=\mysize\textwidth]{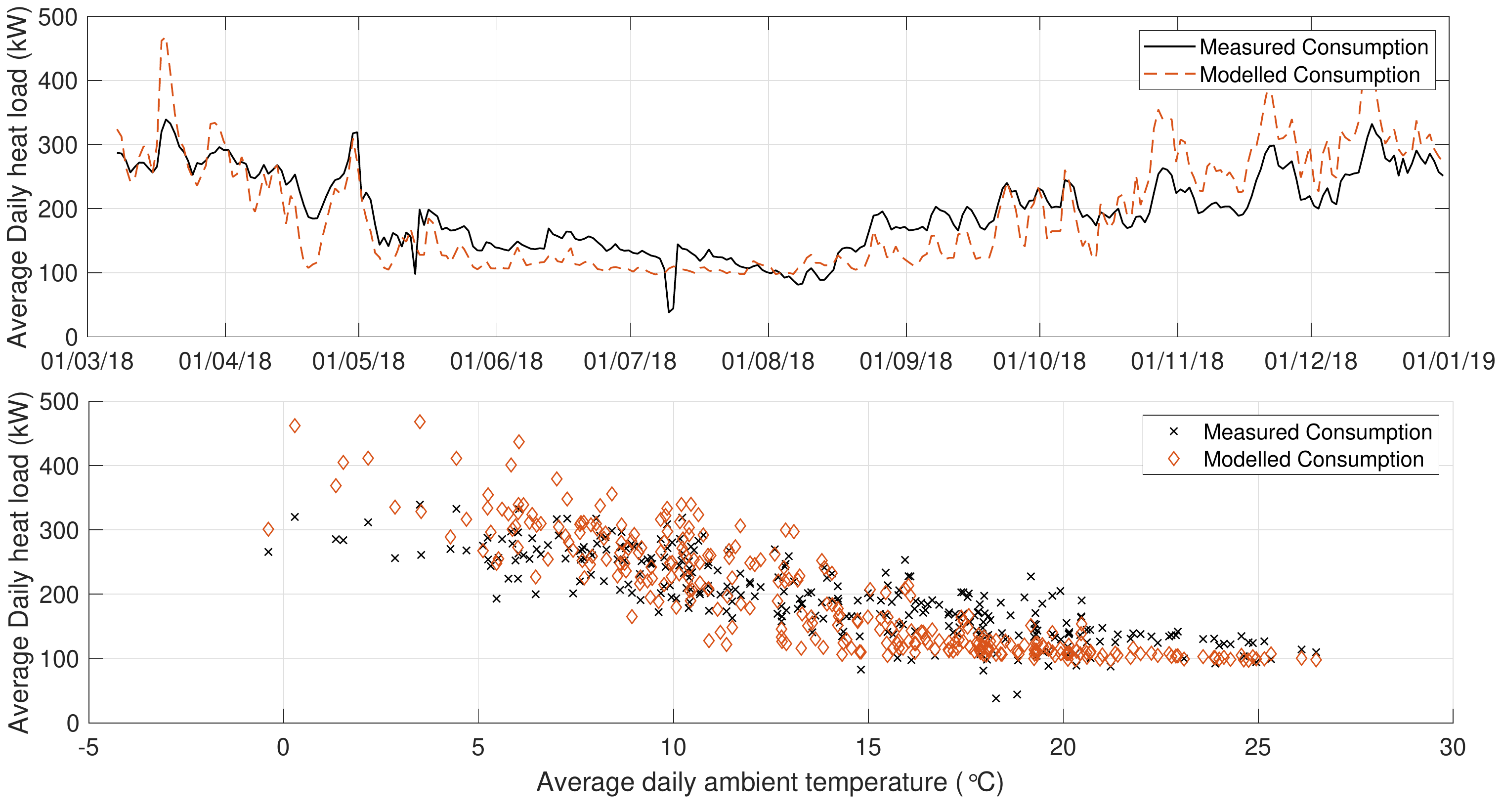}
\caption{\label{fig:Scat}Mean gas consumption vs. mean external air temperature at Ernest Dence: Modelled and measured}
\end{figure*}

A new Water-Source Heat Pump (WSHP) based system is to be installed including heat pumps, boilers and thermal storage. By reducing reliance on gas-based boiler systems and better insulating the buildings, the council seeks to significantly reduce the carbon footprint of the estate. The various components are modelled using generic Ptolemy actors, while the thermal representations of the housing units are developed using EnergyPlus, based on a model initially developed in the MOEEBIUS project \cite{Moeebius}. The heat pump COP (calculated here a function between the source and the sink) is calculated based on the results of \cite{Cameron2013}. To validate the accuracy of the model, the measured pre-retrofit energy consumption over a 10-month period was compared to the predicted energy consumption produced by the model (set up in the pre-retrofit configuration). The plot in Fig.\ref{fig:Scat} shows the measured and modelled average daily energy consumption as well as a scatter plot showing these consumption values plotted against the daily external air temperature. While a certain amount of deviation is expected due to assumptions about the user-behaviour (occupancy hours and internal temperature set-points are not known for example), the modelled consumption profile a good representation of the measured data, particularly capturing the relationship between the external air temperature and the heating requirement.
Using typical building simulation metrics taken from ASHRAE guideline 14 \cite{ASHRAE2002}, the Normalised Mean Bias Error (NMBE) is calculated as 0.7\% while the Coefficient of Variation of the Root Mean Square Error (CV(RMSE)) is 25.9\%. These values are within the ASHRAE guideline targets of $\pm$10\% and 30\% respectively.

The second subsystem captured by the digital twin represents the increased penetration of electric vehicle usage in the district resulting from the installation of additional charging capacity as well as the installation of solar PV panels (with a combined generation capacity of 270 kWp) for power generation.
The EV demand profiles are generated based on a combination of the average charging requirement profiles in the UK taken from \cite{EVData} (300 vehicles are assumed), with an additional stochastic component added to represent real-world uncertainty. It is assumed that 50\% of the EV charging request is associated with car-club and council-owned fleet vehicles, the charging of which can be delayed if required. The conversion of solar radiation to electrical energy is modelled as a function of the area, inclination and tilt of the panels, the latitude of the location, the time of day and year and the direct and diffuse solar irradiance as described in detail in \cite{ODwyer2016}.

\begin{figure*}
\centering
\includegraphics[width=\mysize\textwidth]{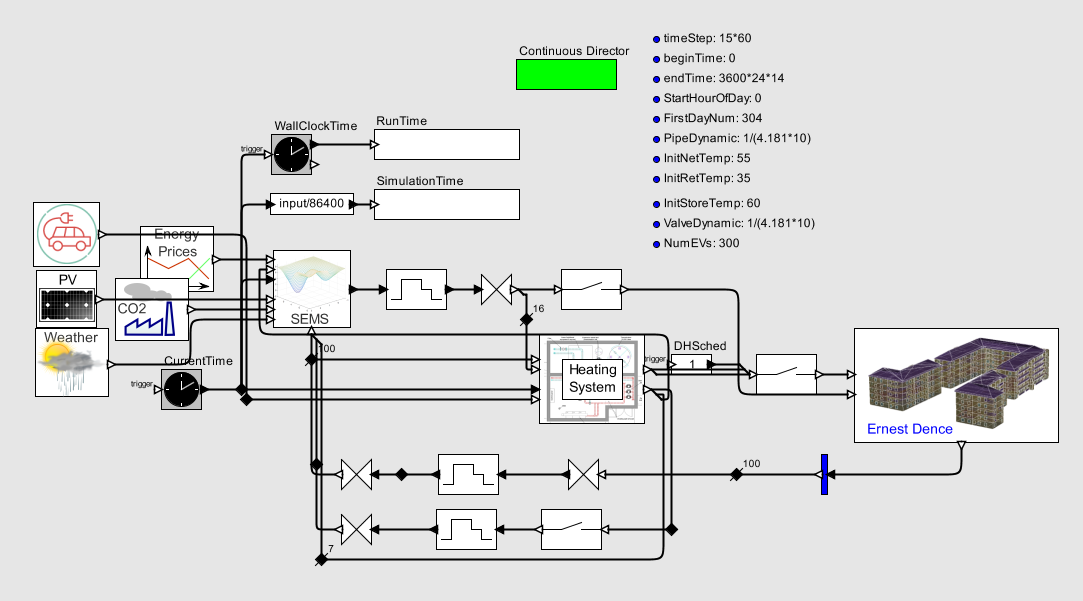}
\caption{\label{fig:TwinInterface}Highest level of Greenwich subsystems digital twin interface}
\end{figure*}

The two subsystems can be simulated concurrently, communicating all relevant sensor measurements at each simulated time-step to the SEMS through a micro web framework (Flask). The SEMS then generates and returns a set of optimised set-points (described in the following sections) for the various simulated energy assets. The highest level of the digital twin implementation in Ptolemy is shown in Fig.\ref{fig:TwinInterface}.

\subsection{Configuration of the SEMS for the Greenwich subsystems}\label{results}
From a decarbonisation perspective, electrification of the heating and transport sectors can offer significant benefits for a district, however several challenges arise with such interventions. For a start, the relatively high cost of electricity may result in increased heating bills for residents, while the likely timing mismatch between solar power generation and EV charger demand will reduce the potential for local utilisation of locally generated power. Furthermore, the new electrical loads will place additional strain on the power network, affecting resilience and potentially leading to a need for significant infrastructural development (which in turn leads to additional costs). The purposes of the SEMS in this scenario are then threefold:
\begin{itemize}
  \item minimisation of some combination of financial cost and environmental impact
  \item maximisation of PV power utilisation for charging of EVs
  \item satisfaction of system-level power constraint, as set by a network operator
\end{itemize}

To illustrate the application of the SEMS to these goals, a set of case studies are presented here using the SEMS with the digital twin, configured as described in Section \ref{GreenwichSetup}. The case studies each demonstrate a 14-day simulation period from the 01/11/2018 - 14/11/2018. External temperature data for London was taken from \cite{WeatherData}, while solar radiation for the area was taken from data presented by the London Air Quality Network \cite{SolarData}. The historical solar radiation with 15 minute time samples was used as an input to the PV generation model - in this way, the stochastic varying nature of generated power is captured. Electricity and gas pricing for the time period was found using the approach described in \cite{Octopus} (with half hourly variations), while the Carbon Intensity API \cite{CarbonAPI} was used to access both predictions and measurements of the $CO_2$ intensity of the grid. A value of 0.184 $kgCO_2/kWh$ was assumed for the $CO_2$ content of natural gas \cite{BEIS2018}. The selection and hyper-parameter tuning of the ML-based forecast generation approaches are not covered in detail in the presented case studies for the sake of brevity.

\subsection{Analysis of performance in heat network application}
The first subsystem (that of the heat network of the Ernest Dence estate) is solely considered here. It is assumed that the SEMS cannot influence the demand side of the heat network (the radiator set-point schedules) and can thus only affect the primary side of the heat network. The network, heat pump and boiler flow temperature set-points and the thermal store charge/discharge set-point are then taken to be the values set by the SEMS for this subsystem. The objective is to minimise a weighted sum of the financial operating cost and the $CO_2$ emissions while maintaining the system temperatures within desired operational bounds. To do this, a state-space model of the heat network is generated which allows for the network temperature to be predicted over a 1-day horizon (with a 15 minute sample time) based on the heat supplied by the supply (heat pumps and boilers) and storage (thermal store) assets and the heat drawn off by the building demand (the Ernest Dence estate). The problem is formulated as an MPC problem, using a quadratic penalty on violations of the heat network temperature bounds and linear penalties on the financial and environmental terms. At each sample, a full day-ahead input trajectory is generated, of which the elements associated with the subsequent time-step are implemented. At the next sample, the initial system states are updated using measurements from the digital twin and the problem is repeated in a receding horizon manner. The problem is implemented in code using the IPOPT solver \cite{Ipopt} called with Pyomo.

\begin{figure*}
\centering
\includegraphics[width=\mysize\textwidth]{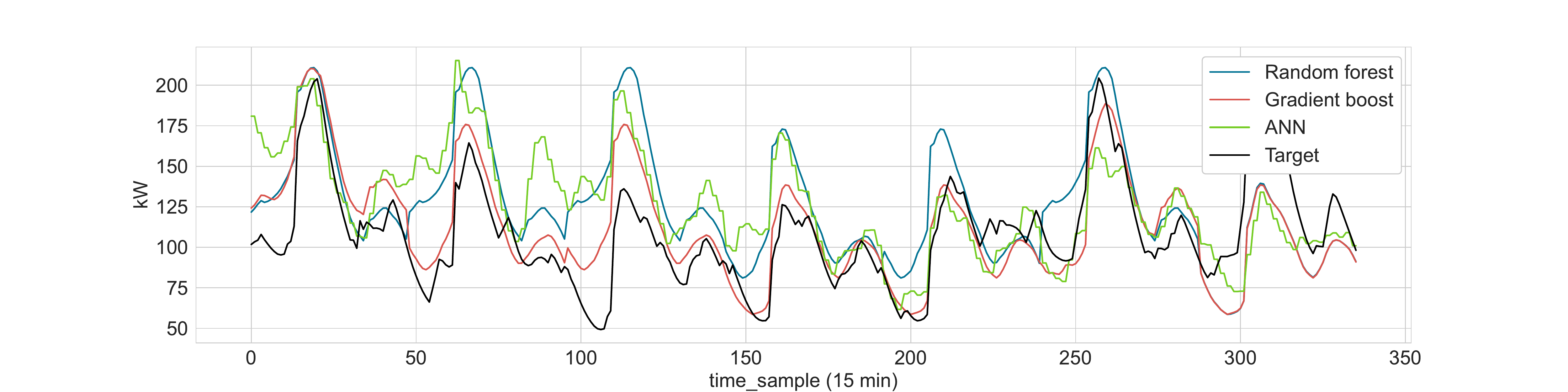}
\caption{\label{fig:MLdemandcomp}ML approaches comparison for forecasting heat demand (one week of validation data shown)}
\end{figure*}

As mentioned, a forecast of the heat demand is required to predict the heat drawn from the heat network. Such a forecast is generated at each sample using the previous 10-months of modelled \textit{historical} consumption data, external temperatures and day types (weekend, holiday, weekday) as training data. The classification-based methodology described in Section \ref{MLForecast} using a gradient-boost approach (with 100 trees and a learning rate of 0.1) is used for the same. The specific ML models as well as their hyper-parameters were chosen through empirical testing using measured data as well as data generated with the digital twin. The methods compared also included a random-forest based classification methods and an ANN-based methods. For illustrative purposes, a comparison of the different tested approaches is shown for a week of validation data in Fig.\ref{fig:MLdemandcomp}. The mean absolute errors for each method calculated over a 60 day unseen validation period are shown in Table \ref{tab:MLdemandcomp} whereby all algorithms were trained using 10 months of data sampled at 15 minute intervals. 

\begin{table*}
\centering
\begin{tabular}{c|c|c|c}
  & Gradient-boost & Random forest & ANN \\
  \hline
  Mean absolute error & 54.7 & 59.1 & 57.4 \\
  \hline
\end{tabular}
\caption{\label{tab:MLdemandcomp}Comparison of heat demand forecasting methods tested using a 60 day unseen data-set}
\end{table*}

\begin{figure*}
\centering
\includegraphics[width=\mysize\textwidth]{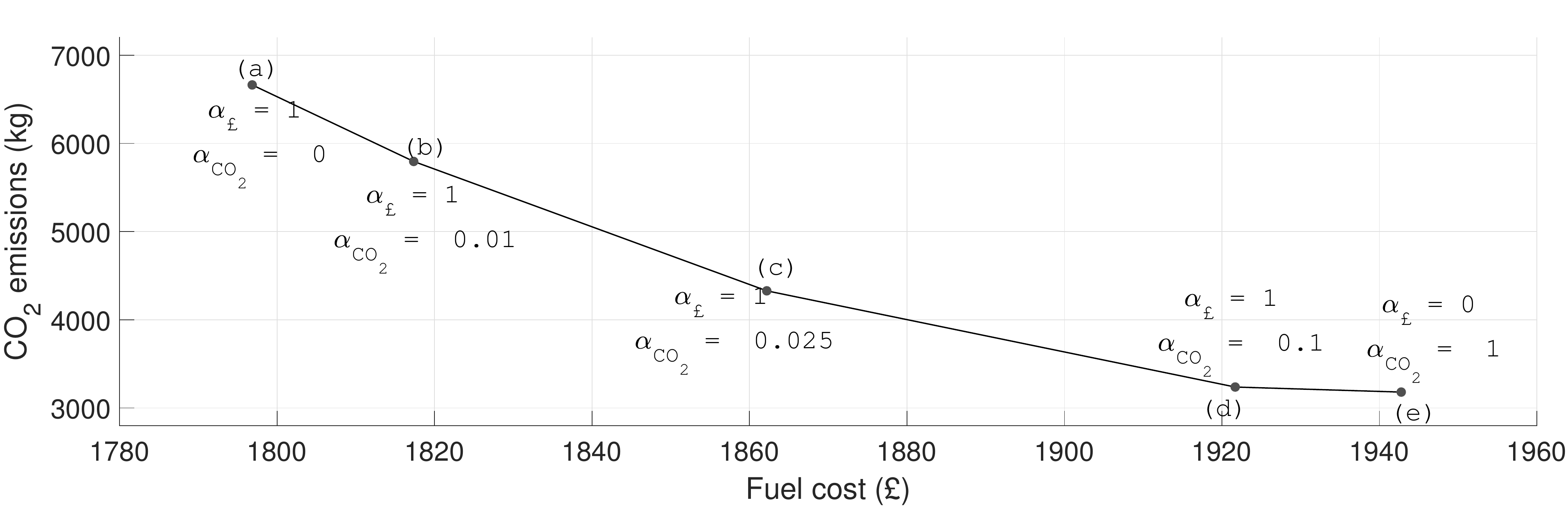}
\caption{\label{fig:Pareto}Cost and environmental performance of the SEMS over 14-day period for different objective weights}
\end{figure*}

The SEMS can raise the network and thermal store temperatures in advance of a future price spike, and switch between boiler-dominated or heat pump-dominated strategies depending on the respective predicted fuel cost and $CO_2$ intensities. There is a clear trade-off, however, between financial and environmental objectives - the use of the boilers always increases the quantity of $CO_2$ emitted (for the data used here), but often represents the cheapest heating option. Choosing the weights that define this balance in the objective function is key to defining the resulting performance of the system. The use of a digital twin offers a useful advantage for decision makers here as more informed decisions can be made by testing several options prior to implementation in the real system. To demonstrate this, the 14-day simulation is repeated 5 times using a different set of objective weights for each. The results of this are shown in Fig. \ref{fig:Pareto} in which $\alpha_{\pounds}$ represents the financial cost weight and $\alpha_{CO_2}$ represents the environmental cost weight. A clear Pareto frontier emerges, enabling a decision maker to choose a desirable trade-off. It can be seen that by spending more, greater emission reductions can be achieved, with diminishing returns.

\begin{figure*}
\centering
\includegraphics[width=\mysize\textwidth]{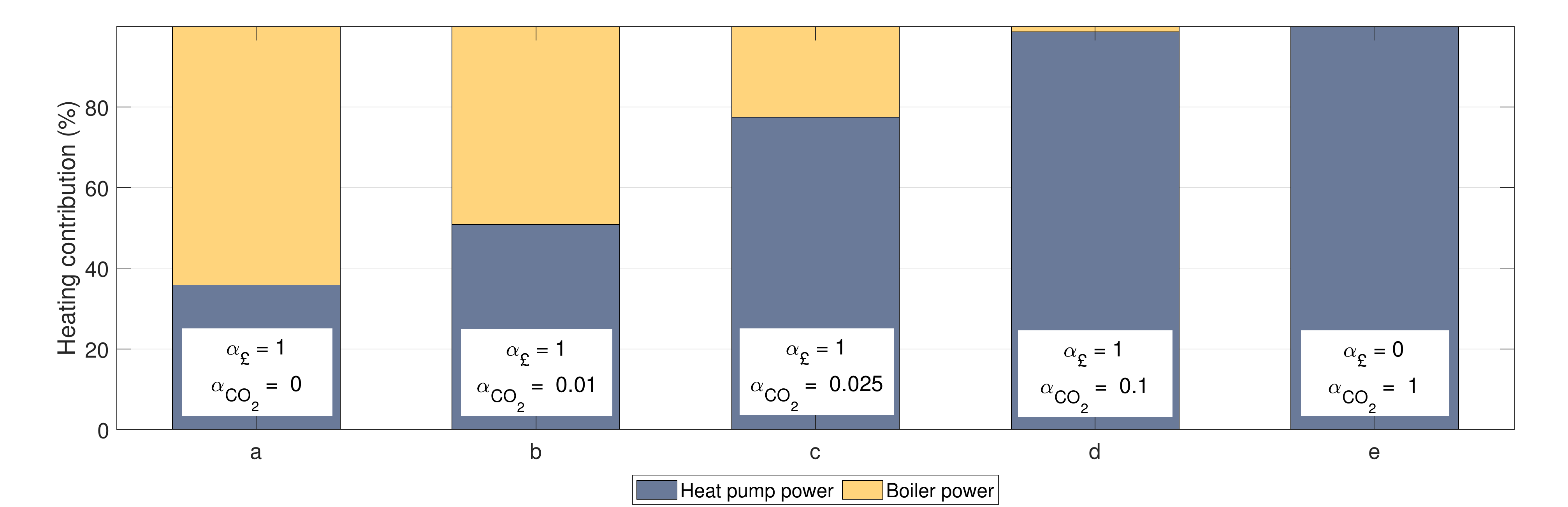}
\caption{\label{fig:Breakdown}Breakdown of heat energy sources for different objective weighting strategies}
\end{figure*}

The heat supply technology mix chosen for the 5 scenarios is shown in Fig.\ref{fig:Breakdown} where labels a-e represent the scenarios from lowest to highest $CO_2$ emission penalties respectively. As expected, a greater emphasis on the environmental objective lead to a reduced reliance on gas. 

\begin{figure*}
\centering
\includegraphics[width=\mysize\textwidth]{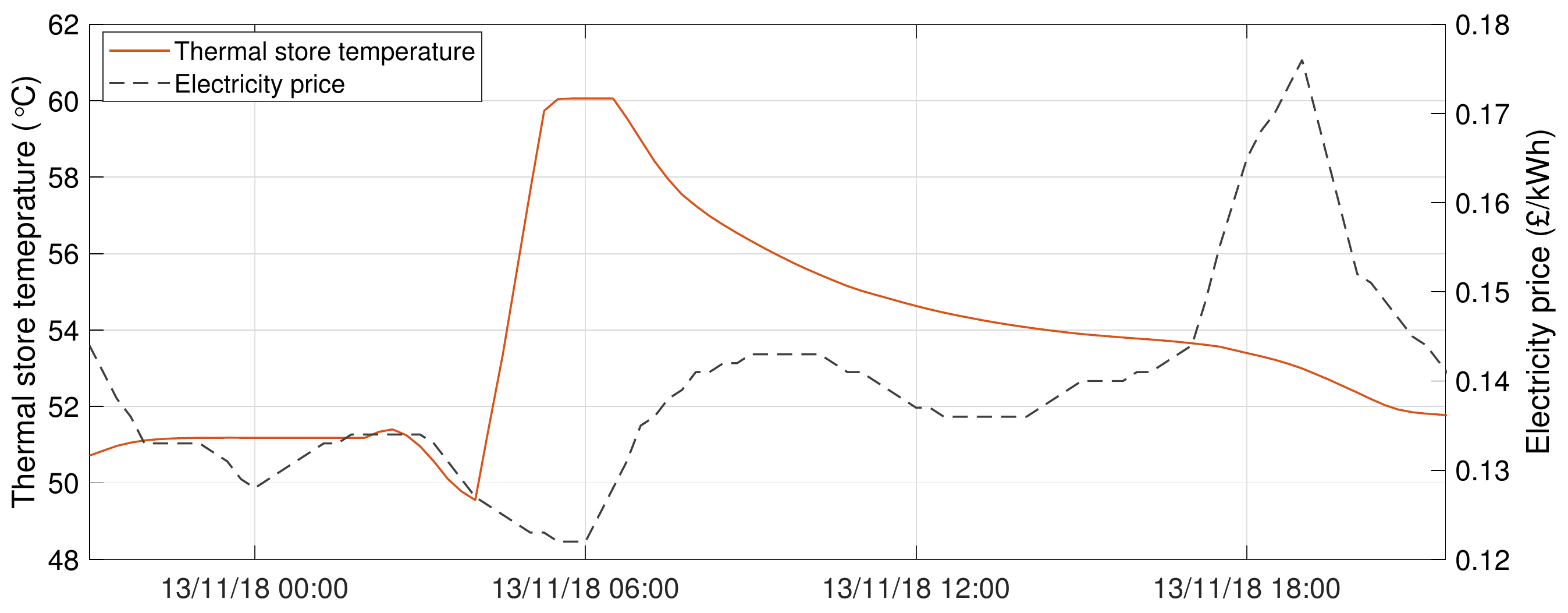}
\caption{\label{fig:TstoPrice}Thermal storage temperature management to minimise fuel cost}
\end{figure*}

The use of the thermal store to facilitate improved performance (in terms of the objective) is illustrated in Fig.\ref{fig:TstoPrice}, where the thermal store temperature profile is shown for one of the simulated days (in scenario $(a)$) along with the electricity price for that day. The store charges to its maximum temperature (60$^{\circ}C$) during a dip in price overnight, before slowly discharging throughout the day when the price is higher. 

\subsection{Analysis of performance in EV/PV application}
In the second subsystem, the SEMS can delay the charging of EVs to maximise the utilisation of the PV generation. For this electrical demand optimisation problem, the power grid and the PV panels are the supply assets - the EV demand can be satisfied by either the generated PV power or the grid. By delaying and reallocating the time of vehicle charging, the electrical demand is reduced and increased at different times in the same manner as would be the case for a battery connected to the system being discharged and charged at different times. Thus, the ability to delay EV demand is characterised here as an electrical storage asset (with a negative storage capacity). The objective of the problem is to minimise the economic cost of EV charging by reallocating \textit{delayable} load in a suitable manner (typically towards times of greater PV production). As mentioned, the delayable load is assumed here to account for 50\% of the demand request. This optimised scheduling is carried out at 15-minute intervals using 1-day ahead forecasts of PV generation, EV demand and electricity pricing information.

\begin{figure*}
\centering
\includegraphics[width=\mysize\textwidth]{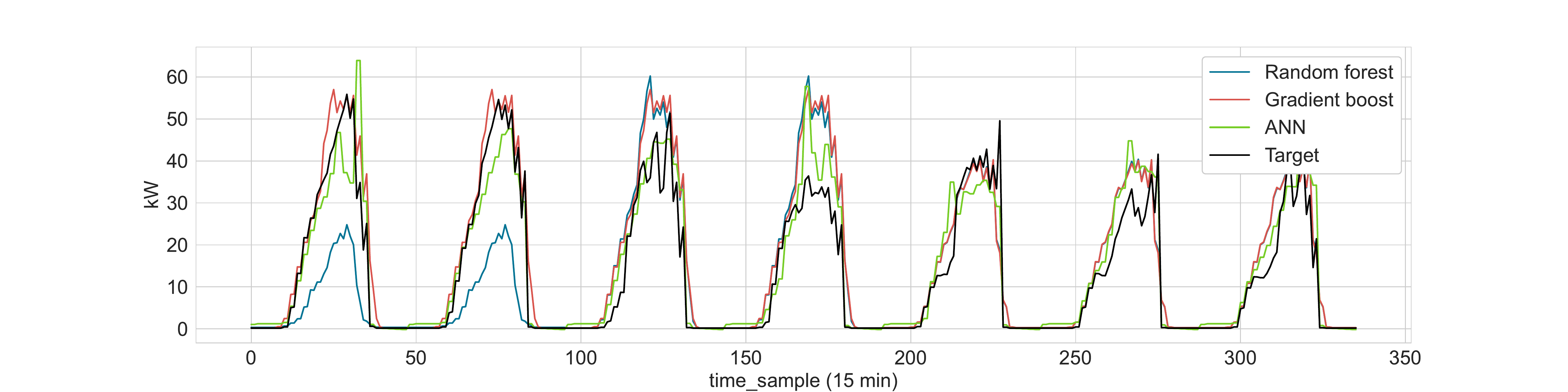}
\caption{\label{fig:MLcomp}ML approaches tested for forecasting PV generation (one week of validation data shown)}
\end{figure*}

To carry out this optimisation, the SEMS must first generate the PV production and EV charger demand forecasts. For the former, the regression-based ANN approach was found to produce the most consistent predictions, while the classification-based gradient boosting approach performed better for the latter. Once again, ANN, gradient-boost and random forset methods were compared. The different tested approaches are shown for a week of validation data in Fig.\ref{fig:MLcomp}. The mean absolute errors for each method calculated over a 60 day unseen validation period are shown in Table \ref{tab:MLPVcomp} whereby all algorithms were trained using 10 months of data sampled at 15 minute intervals. The forecasts are fed to the optimisation problem, formed here as an LP in which the objective function seeks to minimise a combination of the cost of the power drawn from the grid and an additional penalty on the total delayed EV charger power. The latter term is penalised to dissuade the SEMS from suggesting excessive delays. The problem is solved using the GLPK solver \cite{glpk}.

\begin{table*}
\centering
\begin{tabular}{c|c|c|c}
  & Gradient-boost & Random forest & ANN \\
  \hline
  Mean absolute error & 5.2 & 4.8 & 3.9\\
  \hline
\end{tabular}
\caption{\label{tab:MLPVcomp}Comparison of PV forecasting methods tested using a 60 day unseen data-set}
\end{table*}

\begin{figure*}
\centering
\includegraphics[width=\mysize\textwidth]{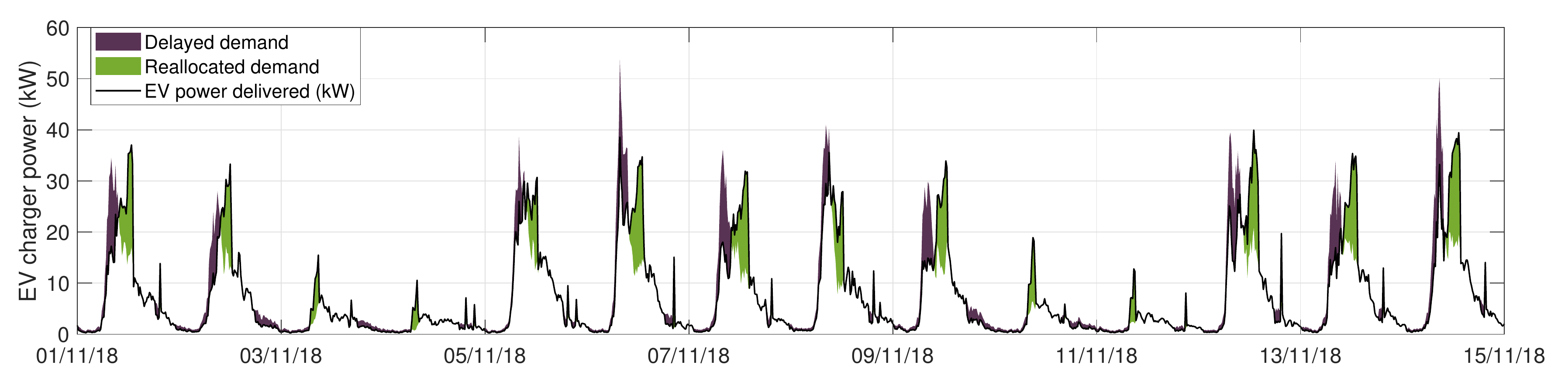}
\caption{\label{fig:EVShift}EV demand profile with reallocation of load to increase PV utilisation}
\end{figure*}

With the SEMS configured in this manner, the 14-day simulation is carried out in the digital twin. A plot of the resulting EV charging power demand is shown in Fig.\ref{fig:EVShift} along with the delayed demand (in grey) and the reallocated demand (in orange). It can be seen that on most days, the SEMS delays some of the peak EV demand and reallocates it to a later point in the day to more closely coincide with the solar generation peak.

The outcome of this demand reallocation is summarised in Table \ref{tab:EVShift} in which different outcomes are shown for a case with and without the SEMS delay/reallocation. It can be seen that the percentage of the generated PV that is utilised by the EV-chargers is increased, along with the renewable share of the EV charger consumption. The cost savings and $CO_2$ savings resulting from this reduction in grid consumption is also shown for the 14-day period.

\begin{table*}
\centering
\begin{tabular}{c|c|c|c|c}
  & \% PV & \% PV & Cost & $CO_2$\\
  & Utilisation & Contribution & (£) & (kg)\\
  \hline
  No delay & 54.0 & 45.2 & 746 & 1339\\
  Delay & 60.7 & 50.8 & 670 & 1206\\\hline
\end{tabular}
\caption{\label{tab:EVShift}Impact of delaying EV charger request to increase PV utilisation}
\end{table*}

\subsection{Coordinated operation}
The two subsystems have quite a different focus, however, in both cases the electrical power network is a key component. New heat pump and EV assets in the district may lead to increased stress on the grid, particularly in times when the demand peaks coincide. A mechanism whereby a total electrical power supply is limited by the SEMS would prevent such occurrences, increasing resilience and lessening the need for potentially significant grid reinforcement. To illustrate the ability of the SEMS in this regard, a final case study is presented here whereby a higher-level coordination layer can check if a pre-defined supply limit is exceeded and provide alternative subsystem solutions that eliminate the violation.

The previous 2 subsystems are simulated again for this case study, with the heating-based subsystem configured to replicate scenario $(c)$ and with EV demand reallocation enabled in the EV/PV subsystem. A limit of 70kW is set in the coordination layer on the total grid power available to the heat pumps and the EVs. When the SEMS predicts a violation of this limit at any point in the prediction horizon, the electrical supply limit to the heating-based subsystem is limited in the manner described in Algorithm \ref{Alg} and the optimisation of this subsystem is repeated. 

\begin{figure*}
\centering
\includegraphics[width=\mysize\textwidth]{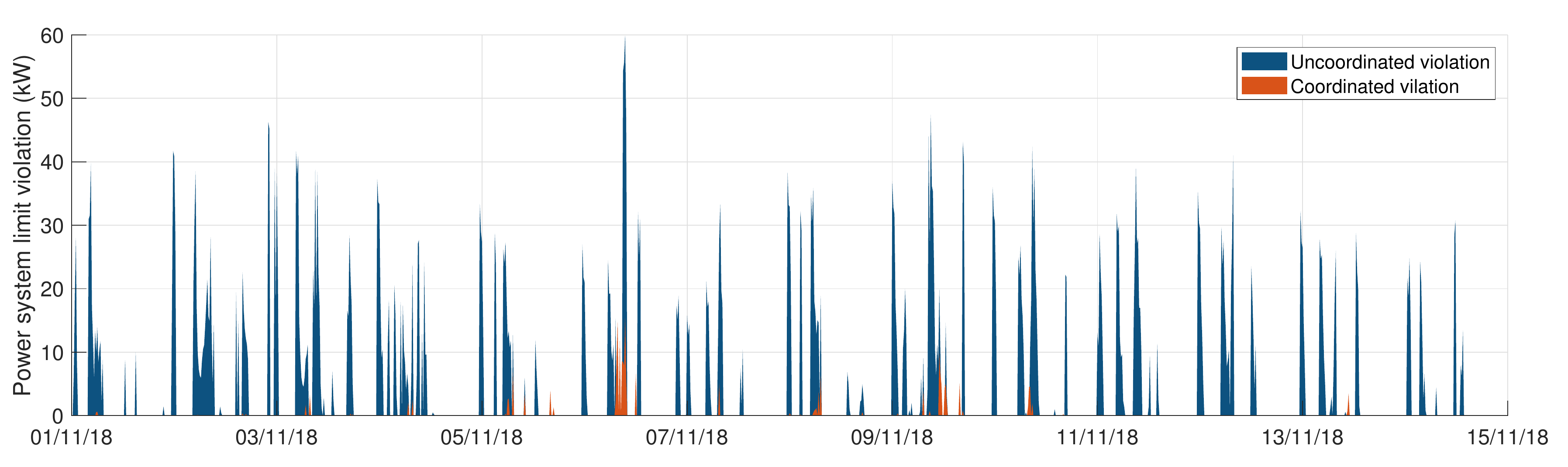}
\caption{\label{fig:Viol}Violation in district-level power supply constraint with and without master-coordinator}
\end{figure*}

The magnitudes of the violations of the system limit with and without this coordination mechanism are shown for the 14 day simulation in Fig.\ref{fig:Viol}. It can be seen that violations of the limit are significantly reduced when the SEMS can coordinate the subsystems in this manner with the exception of some minor violations which occur as a result of mismatches between the predicted and measured EV demand and PV generation supply.

As a quantifiable metric, the sum of the magnitudes of these violations were compared. Under coordinated management, the violations were reduced by approximately 97\%. Despite the same demand energy requirement, the imposition of this constraint led to an increase of 20.8\% in $CO_2$ emissions and 8.9\% in energy cost for the first subsystem due to the additional constraints on the system (leading to additional boiler load). Once again, the usefulness of the digital twin should be noted here - in this case the ability to use the digital twin to quantify the expected impact of such a limiting constraint. Such insight can inform infrastructural investment decisions, quantify risk scenarios and devise possible mitigation strategies.

\section{Conclusions}
The transition to more interconnected urban energy landscapes leads to challenges in terms of stresses to the current energy infrastructure and opportunities in terms of the application of intelligent control and coordination strategies. To leverage the latter opportunity to overcome the former challenge, this paper presents an energy management tool for control and coordination of interconnected energy assets in a smart city or district referred to as the SEMS. A modular framework is introduced in which ML-based forecast generation and grey-box model fitting algorithms are used to generate predictive components used within a flexible MPC-based control strategy. The strategy allows for subsystems to be considered independently with an additional coordination layer used to impose higher district-level constraints and limitations. A simulation framework is presented in parallel to act as a digital surrogate of a multi-vector energy system, thus enabling the design, tuning, testing and evaluation requirements prior to deployment of the energy management strategy.

Using the interventions currently being carried out by Greenwich as part of the Sharing Cities project as a basis, a set of case studies were examined to illustrate the application of the SEMS. The potential for balancing environmental and financial objectives in a hybrid building heating system was shown, indicating the trade-offs that must be considered by decision makers in a smart city. A further case study was used to show the potential of the SEMS for increasing local utilisation of renewable energy in an urban context (incorporating EV charging loads). A final example was then used to demonstrate the coordinating capabilities of the SEMS. Incorporating these predictive, modelling, control and coordination aspects in a single flexible tool and ensuring an open architecture for integrating subsequent algorithms and energy sub-systems in the framework, paves the way for a quicker transition to the low carbon economy goals. 

\section{Acknowledgements}
This project has received funding from the European Union's Horizon 2020 research and innovation programme under Grant Agreement No. 691895 and the EPSRC (Engineering and Physical Sciences) under the Active Building Centre project (reference number: EP/S016627/1). The authors would also like to thank Siemens UK, the Greater London Authority and the Royal Borough of Greenwich with whom collaboration on the project has been carried out. IP would also like to acknowledge the Imperial College Research Fellowship scheme for supporting him.


\end{document}